%% file: main.tex
\documentclass{AUJarticle}
\usepackage[cmex10]{amsmath}
\usepackage[nocompress]{cite}
\usepackage{graphicx, multirow, booktabs, color, setspace}
\usepackage{balance}
\usepackage{comment}
\usepackage[hidelinks=true]{hyperref}
\graphicspath{{figures/}}

\begin{document}

\title{Toward Federated Cognitive Digital Twins over the Edge–to-Cloud Continuum}

\addauthor{Alessandra Somma}
{University of Naples Federico II, Italy} 
  {alessandra.somma@unina.it}

\addauthor{Alessio Bucaioni}
{M\"alardalen University, Sweden} 
  {alessio.bucaioni@mdu.se}

\issuev{45}
\issuen{1}
\issued{June 2026}

\shortauthor{A. Somma and A. Bucaioni}
\shorttitle{Toward Federated Cognitive DTs over the Edge-to-Cloud Continuum}

\thispagestyle{plain}

\maketitle

\begin{abstract}
\small
Digital Twins (DTs) are increasingly adopted to monitor, analyze, and optimize Cyber-Physical Systems (CPSs) through continuous interaction between physical assets and their digital counterparts. However, current DT architectures often rely on centralized and monolithic designs, leading to scalability, latency, and resilience issues in distributed environments such as smart cities. Moreover, they provide limited support for semantic integration and high-level reasoning, reducing the effectiveness of DT-based decision-making.

Recent studies on Federated Digital Twins (FDTs) have addressed scalability by decomposing complex systems into interacting twins, but they still largely centralize intelligence in cloud components. In parallel, Cognitive Digital Twins (CDTs) enhance DTs with semantic reasoning, explainability, and AI-driven decision support, yet they are typically difficult to integrate into distributed architectures.

This paper proposes a \textbf{Federated Cognitive Digital Twin} (FCDT) architecture that combines federation and cognition within a unified approach. The architecture distributes intelligence across the edge-to-cloud continuum through local twins, which provide real-time monitoring and lightweight cognitive capabilities, and global twins, which perform system-level reasoning, simulation, and coordination. By integrating distributed autonomy with cognitive reasoning, the proposed approach improves scalability, responsiveness, and decision-making in complex distributed CPSs.


\end{abstract}

\input{sections/01_introduction}

\input{sections/02_related}

\input{sections/03_proposal}
\input{sections/04_conclusion}

\section*{Acknowledgment}
\label{sec:ack}
This work is supported by: (a) the Italian National Center on HPC, Big Data and Quantum Computing through the project ``ECHO-TWIN: Edge-Cloud-HPC Optimized Twins based on Workflow-enhanced Inference Networks'', under the MUR Call Decreto Direttoriale n. 307, 18/03/2025, PN RIC 2021--2027, MI 1\_91449080372\_0000701, CUP: B69H26001070007, B62F26000630005, B69J26002770005; (b) the project "iSecure: Developing Predictable and Secure IoT for Autonomous Systems" (2023-01899); (c) by the Key Digital Technologies Joint Undertaking through the project "MATISSE: Model-based engineering of digital twins for early verification and validation of industrial systems" (101140216); (d) and by the Clean Energy Transition Partnership through the project ``FLEXI: Human-centered AI and digital twin powered energy system integration for flexibility markets" (101069750). 
\bibliographystyle{ieeetr}
\bibliography{bibliography}
\balance

\end{document}

%% file: sections/01_introduction.tex
\section{Introduction}
\label{sec:introduction}
Digital Twins (DTs) are virtual representations of Cyber-Physical Systems (CPSs), characterized by continuous bidirectional communication between the physical and digital domains \cite{Ferko_Architecting_2022}. This interaction enables the transfer of real-world data from the physical system to its digital counterpart, while also allowing the DT to generate insights and provide feedback to the physical world \cite{DeBenedictis_DTHealth_2023, Tao_DTModeling_2022}. By combining data, models, and analytics, DTs support advanced monitoring, enhance situational awareness, and enable data-driven decision-making across domains such as manufacturing, transportation, and smart cities \cite{bucaioni2025multi, somma2025model}.

Engineering and architecting DT is a complex and demanding activity, and for this reason it has received growing attention in recent years. This interest has led to the proposal of many DT architectures \cite{Ferko_Architecting_2022, Minera_DTIoT_2020}, together with important standardization efforts \cite{ferko2023standardisation, ferko2012supporting}. However, when DTs are applied to large-scale and inherently distributed environments such as smart cities, new challenges emerge \cite{somma2025model}. In these contexts, heterogeneous devices continuously generate data, multiple subsystems interact with one another, and decisions often need to be taken in a timely and coordinated way. As a result, current DT approaches reveal two major weaknesses: they are still largely based on centralized designs \cite{Ferko_Architecting_2022, Duran_DTaaS_2026, SOMMA2026_twinarch}, and they provide limited support for semantic integration across heterogeneous subsystems and information sources \cite{ferko2023analysing, SOMMA2026_twinarch}.

Most existing DT architectures address this complexity through centralized solutions. 
In centralised designs, the DT is usually implemented as a monolithic system, often deployed in the cloud, and designed as a single unit representing the whole real system. 
At the same time, data generated by distributed physical assets are transmitted to centralized platforms, where are then stored and processed to feed simulation models, analytics pipelines, and AI-based inference engines. 
This design has clear advantages: it simplifies DT management, supports large-scale data processing, and enables advanced monitoring and simulation capabilities. However, these benefits come at a cost. The concentration of both the DT logic and the data processing pipeline in centralized infrastructures introduces significant challenges in terms of latency, bandwidth consumption, and resilience \cite{Shi_EdgeComputing_2016, Satyanarayanan_EdgeComputing_2017}. Centralized solutions remain also sensitive to network disruptions and make local autonomy difficult to achieve \cite{Pasupulati_CentralizedArchitectures_2016}.


For these reasons, DT research is progressively moving beyond monolithic and fully centralized designs toward more distributed solutions. In this direction, the concept of \textbf{Federated Digital Twins} (FDTs) has emerged as a promising way to decompose the complexity of large CPSs \cite{Vergara_FDT_2023, Vergara_FDT_2024}. The main idea is to model the overall system as a set of autonomous yet interoperable twin instances, each associated with a specific subsystem, while allowing them to cooperate in order to provide a coherent system-level view \cite{Vergara_FDT_2024, Papacharalampopoulos_FDT_2024}. In this way, federation addresses an important structural limitation of monolithic DTs, since the complexity of the global system is no longer handled by a single twin, but distributed across multiple coordinated twins.

However, the FDT paradigm is still in an early stage, and many existing proposals remain at a rather high level of abstraction. In particular, they often lack a clear architectural framework to support coordinated intelligence across distributed twins \cite{Michael_MDE4DT_2025, somma2025model, Vergara_ApproachFDT_2025}. Moreover, even when federation is adopted, intelligence is frequently still concentrated in cloud-based components. This helps in managing the overall complexity of the DT, but it does not fully solve the problems that motivated the shift away from centralization in the first place, especially latency, resilience, and local autonomy. The result is that currently federation of DTs improves their structural decomposition, but often without yet providing a fully effective distribution of their intelligence \cite{Michael_MDE4DT_2025, Vergara_ApproachFDT_2025}.

Hence, the shift toward federation helps in addressing the structural complexity of distributed CPSs \cite{Manzi_interTwin_2026, Vergara_FDT_2024}, but it does not fully solve their limited semantic capabilities. 
As a consequence, DTs are often able to detect or predict events, but they struggle to explain them, relate them across subsystems, or support reasoning about their implications. For example, identifying a delay in a transportation network does not automatically provide an understanding of its causes, how it propagates through the system, or which actions should be taken to mitigate its impact. This gap between data processing and meaningful reasoning limits the effectiveness of DTs as decision-support systems \cite{Liu_CDT_2025}.

To address this limitation, recent research has introduced the concept of \textbf{Cognitive Digital Twins} (CDTs) \cite{Liu_CDT_2025, Zheng_EmergenceCDT_2022}, which extend traditional DTs with capabilities for semantic reasoning, knowledge integration, and explainability. The idea is to move from DTs that only process data to DTs that can also interpret it, relate it to domain knowledge, and support human understanding \cite{Zhang_EngineeringCDT_2020, Abburu_CogniTwin_2020}. In this context, advances in Artificial Intelligence (AI), and in particular in Generative AI (GenAI), play a key role. Large Language Models (LLMs) have shown strong capabilities in integrating heterogeneous information, reasoning across domains, and interacting with users through natural language \cite{Liu_CDT_2025, Xia_DTLLMArchitecture_2025, Zhu_CDTLLM_2026}. 

However, introducing this type of intelligence also raises new challenges. LLMs typically require significant computational resources and are commonly deployed in centralized environments. This creates a tension with the need for distributed DT architectures, especially in scenarios where low latency and local autonomy are required \cite{Singh_LLMFederation_2026}. If not carefully designed, the integration of such models risks reinforcing centralization, which is precisely one of the limitations that federated DTs aim to overcome. 

Recent developments suggest that this tension can be mitigated by distributing intelligence across the edge–to-cloud continuum \cite{Friha_LLMEdge_2024, Cai_EdgeLLM_2024, Hao_SLM_LLM_2024, Rosendo_EdgeCloudContinuum_2022}. In this setting, lightweight AI models, including distilled or specialized versions of language models, can be deployed at the edge to support local tasks such as filtering, summarization, and contextual interpretation \cite{Rao_ECOLLM_2024}. At the same time, more computationally intensive models can be deployed in the cloud to provide global reasoning, explanation, and decision support \cite{Cai_EdgeLLM_2024, Friha_LLMEdge_2024}. 

In this paper, we propose to address these challenges by proposing a unified architecture combining federation and cognition, resulting in the \textbf{Federated Cognitive Digital Twin} (FCDT) architecture. 
In this architecture, local twins provide fast and autonomous decision-making close to the physical system, while global twins support system-level reasoning and coordination. By combining these two levels, the proposed approach maintains responsiveness while enabling advanced analysis and explanation, and at the same time simplifies the engineering of complex DT systems and mitigates latency issues in distributed environments.

%% file: sections/02_related.tex
\section{Related Work}
\label{sec:related}
Existing research on engineering and architecting DTs has evolved along multiple directions, addressing different aspects of DTs design, distribution, and intelligence. In this section, we focus on two lines of research, which are relavant to our work, i.e., Federated DTs and Cognitive DTs.

Early DT architectures are predominantly based on centralized designs, where data from distributed assets are collected and processed in cloud environments \cite{Ferko_Architecting_2022, Minera_DTIoT_2020, Duran_DTaaS_2026, SOMMA2026_twinarch}. FDTs have been introduced to address the structural limitations of monolithic DT architectures in large-scale systems. Instead of representing the entire real system through a single DT instance, FDT approaches decompose it into multiple interacting twins, each associated with a specific subsystem.

Vergara \textit{et al.} \cite{Vergara_FDT_2023, Vergara_FDT_2024} define FDTs as a distributed set of twin instances that collaborate to support system-level decision-making. Their work emphasizes the role of federation in enabling collaborative simulation and coordination, where different twins exchange information to improve global understanding. 
Papacharalampopoulos \textit{et al.} \cite{Papacharalampopoulos_FDT_2024} analyze federation from a modeling perspective, highlighting limitations in knowledge transfer and interoperability between twin instances. Their work shows that even when multiple twins are connected, inconsistencies and lack of shared semantics can hinder effective collaboration. This reinforces the need for stronger integration mechanisms beyond simple federation.

Model-driven approaches, such as the one proposed by Michael \textit{et al.} \cite{Michael_MDE4DT_2025}, investigate how FDTs can be engineered using abstraction and modeling techniques. These approaches contribute to structuring the development process, but they still operate at a high level and do not explicitly address runtime coordination or distributed decision-making.

More recent efforts, such as \cite{Vergara_ApproachFDT_2025}, explore learning-based coordination across federated twins, while large-scale initiatives like interTwin \cite{Manzi_interTwin_2026} combine federation with distributed computing infrastructures. These works show the potential of FDTs in complex environments, but they also reveal a common limitation: intelligence is often still centralized, typically in cloud-based components that aggregate and process data from all twins.

As a result, current FDT approaches successfully address structural decomposition and scalability, but only partially address distributed intelligence. In many cases, federation is used to organize the system, but decision-making and reasoning remain centralized, which limits responsiveness and local autonomy.

In parallel, CDTs have been proposed to address the limited reasoning capabilities of traditional DTs. CDTs extend DTs with semantic reasoning, knowledge integration, and explainability, moving beyond purely data-driven analytics \cite{Zheng_EmergenceCDT_2022, Liu_CDT_2025}.  Zheng \textit{et al.} \cite{Zheng_EmergenceCDT_2022} introduce the concept of cognitive DTs as systems that integrate data, models, and knowledge to support decision-making. Their work highlights the importance of combining physical models with higher-level reasoning mechanisms, but does not provide a concrete architectural realization. 

Early CDTs works such as \cite{Abburu_CogniTwin_2020} and \cite{Zhang_EngineeringCDT_2020} propose hybrid DT architectures that combine data-driven models with symbolic reasoning and self-awareness mechanisms. These approaches represent an important step toward cognitive DTs, but their scope is often limited to specific domains and they do not address large-scale distributed systems.

Liu \textit{et al.} \cite{Liu_CDT_2025} present a comprehensive survey on CDTs and discuss how AI techniques can enhance DT capabilities. In particular, they identify the role of knowledge graphs, reasoning engines, and more recently LLMs in enabling semantic understanding and interaction. Indeed, with the rise of Generative AI, several works have explored the integration of LLMs into DT systems. For instance, Xia \textit{et al. } \cite{Xia_DTLLMArchitecture_2025} propose an architecture where LLMs are used to interact with DTs and support reasoning over system data. Similarly, Zhu \textit{et al.} \cite{Zhu_CDTLLM_2026} discuss how semantic communication and LLM-based reasoning can enhance DT systems. These works demonstrate the potential of LLMs in enabling explainability, interaction, and cross-domain reasoning.


Recent research has started to explore how AI models, including LLMs, can be distributed across the edge–to-cloud continuum. Surveys such as \cite{Friha_LLMEdge_2024} highlight the potential of deploying lightweight models at the edge to support local processing, while keeping more complex models in the cloud.

Frameworks such as Edge-LLM \cite{Cai_EdgeLLM_2024} and hybrid inference approaches \cite{Hao_SLM_LLM_2024} propose splitting LLM execution across edge and cloud resources. These approaches enable tasks such as filtering, summarization, and contextual interpretation to be performed locally, reducing communication overhead and improving responsiveness. Similarly, \cite{Rao_ECOLLM_2024} explores optimization strategies for edge–cloud LLM deployment.

These approaches suggest that LLM-based capabilities can be adapted to distributed environments through techniques such as model distillation, specialization, and collaborative inference. In this setting, edge components handle lightweight tasks, while cloud components perform more complex reasoning and decision-making \cite{Rao_ECOLLM_2024}.  Nevertheless, even though these works provide important insights into distributed AI, they are not specifically designed for DT systems. In particular, they do not define how distributed AI components should be integrated within a DT architecture, nor how they should interact with local and global twin instances.

\smallskip
The analysis above highlights a clear gap in the current DTs research. Federated approaches address the decomposition of complex systems into multiple twins, but often lack mechanisms for distributing intelligence and coordinating decisions across them. Cognitive approaches enhance reasoning capabilities, but are typically centralized and do not account for the constraints of distributed CPSs. Research on edge AI and LLM distribution provides enabling technologies, but does not offer a DT-specific architectural framework. As a result, existing solutions tend to address either distribution or cognition, but rarely both in a unified way. This gap motivates the need for a new approach that combines federation and cognition within a single architecture.

%% file: sections/03_proposal.tex
\section{Federated Cognitive Digital Twin Architecture}
\label{sec:proposal}

\begin{figure*}[!ht]
	\centering
	\includegraphics[width=0.6\textwidth]{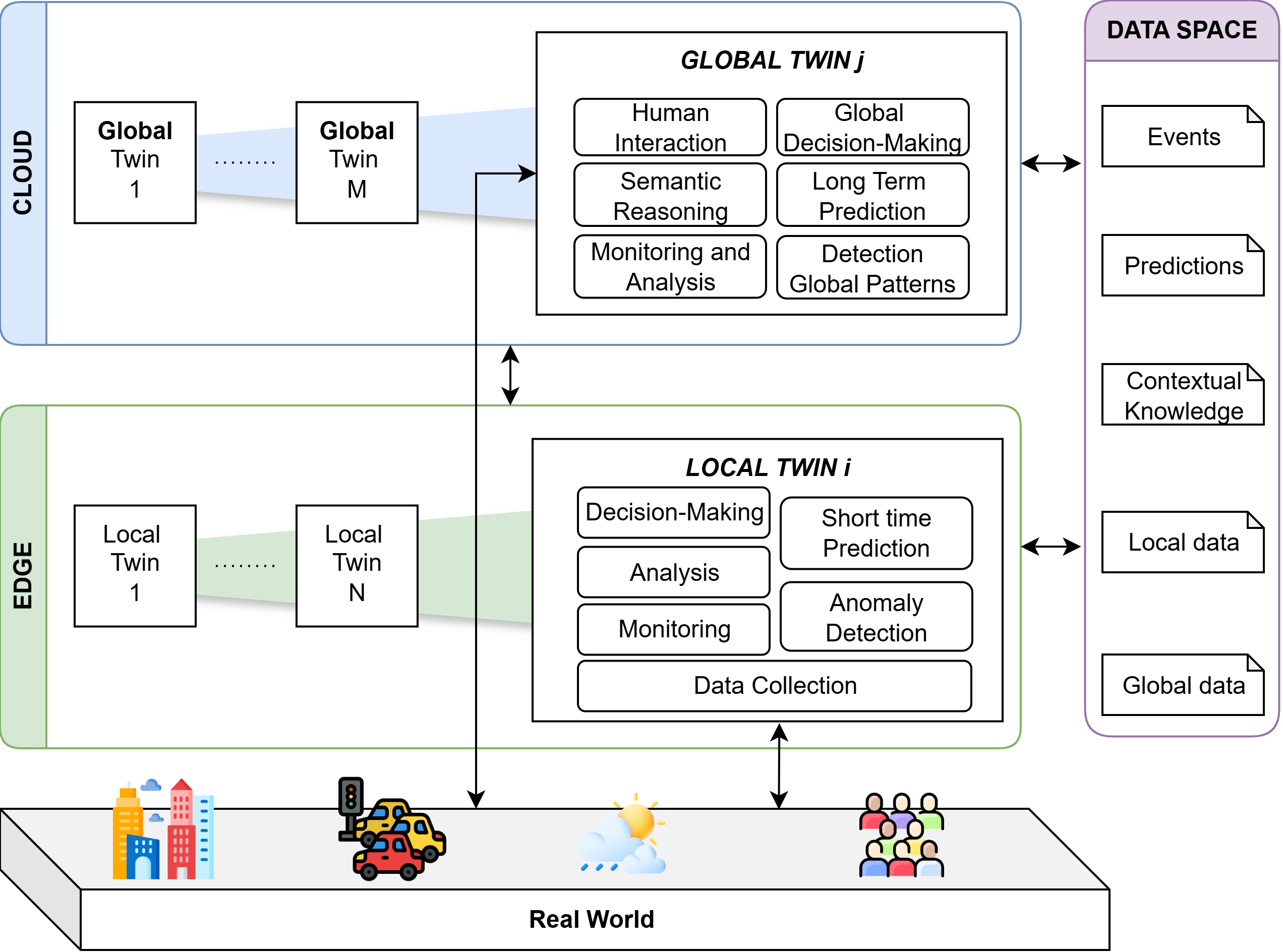}  
	\caption{Federated Cognitive Digital Twin Architecture.}
    \label{fig:fcdt}
\end{figure*}
The proposed FCDT architecture models the Digital Twin as a federated and cognitive system distributed across the edge-to-cloud continuum. Figure \ref{fig:fcdt} presents the overall architecture, which distributes DT capabilities across the continuum through two complementary types of twins: \textit{local twins} and \textit{global twins}. Local twins are associated with specific physical subsystems and deployed at the edge, close to the physical environment, while global twins operate at the cloud level and reason over the information generated by multiple local twins.

The key principle of the architecture is bringing part of the DT capabilities closer to the real world system. This is not only a deployment choice, but a functional one. Local twins are not limited to collecting and forwarding data; they provide monitoring, analysis, and a first level of decision-making directly at the subsystem level. This allows them to maintain tighter synchronization with real-world conditions and to react quickly to local changes. The cloud remains essential, but its role shifts from being the central point of intelligence to the level where information are aggregated, correlated, and used for broader reasoning, simulation, and coordination.

Each local twin maintains a continuously updated representation of its subsystem by ingesting data from sensors, devices, and local services. This state forms the basis for real-time analysis, including anomaly detection, short-horizon prediction, and state estimation. Since these activities are performed close to the data source, they can be executed with low latency and without relying on continuous cloud communication, which is crucial in dynamic and time-critical environments.

Based on this analysis, local twins support a first level of decision-making. By combining current state and analytical results, they can take immediate actions within the scope of the subsystem they are twinning. Although these decisions are inherently local, they provide fast and autonomous responses that would be difficult to achieve in centralized systems. In this way, local twins contribute directly to system responsiveness and resilience.

In addition to these operational capabilities, local twins incorporate lightweight cognitive components, including edge AI and specialized or distilled edge language models. These components support semantic processing tasks such as event extraction, summarization, and contextual interpretation. This enhances the quality of local decisions by enriching numerical data with contextual meaning, which can be useful for human operators or local control logic. At the same time, it enables local twins to transform raw data into structured and compact representations before sharing it. This process, referred to as semantic distillation, reduces communication overhead while preserving relevant information. It can also support privacy-aware data sharing by enabling the transmission of abstracted or anonymized information.

As a result, local twins play a dual role. They act as operational components, responsible for real-time monitoring and local decisions, and as cognitive filters, responsible for transforming raw observations into structured knowledge. These outputs are stored in a shared data space and constitute the input for higher-level reasoning.

Global twins operate at the cloud level and provide a system-wide perspective. Their role is to reason over the structured information produced by multiple local twins and other global twins, and derive knowledge that cannot be obtained from a single subsystem. Unlike local twins, they do not directly interact with the physical environment, but consume data from the shared data space in the form of events, predictions, and contextualized observations. This abstraction allows them to focus on higher-level and more computationally demanding tasks.

Using these aggregated information, global twins perform system-level analysis, including the identification of correlations across subsystems, detection of global patterns, and long-term prediction. They are also responsible for simulation and the exploration of alternative scenarios. By leveraging the computational resources of the cloud, they can execute complex and time-consuming analyses, such as what-if simulations and the evaluation of coordinated strategies, supporting more informed and anticipatory decision-making.

Global twins are also the main location where the architecture realizes its full cognitive capabilities. They can exploit more powerful AI models, including full-scale LLMs, to support semantic reasoning, explanation, and human interaction. At this level, LLMs are used to interpret aggregated information, relate observations across subsystems, explain system behavior, and assist in understanding the impact of potential actions. In this way, global cognition complements local cognition, providing deeper and broader reasoning over the entire system.

The interaction between local and global twins is mediated by a shared \textit{data space}, which is depicted in vertical in Fig. \ref{fig:fcdt} and acts as the integration layer of the architecture. This component enables information exchange and coordination without tightly coupling the twins. Instead of direct communication, all components interact through a common information space where structured data, events, summaries, and inferred knowledge are stored and accessed. This design supports decoupling and scalability, as twins do not need to be aware of each other's internal structure.

Through this mechanism, local twins continuously generate structured information and propagate it to the data space. Global twins consume this information, reason over the system state, and produce higher-level insights, strategies, or recommendations. These can then be propagated back to the edge, where local twins use them to refine their behavior. This establishes a bidirectional flow of information, combining bottom-up knowledge propagation with top-down coordination.

Overall, the proposed architecture defines a distributed organization of intelligence. Instead of concentrating all capabilities in the cloud, DT intelligence is explicitly divided across levels according to their role and constraints. At the edge, DT intelligence is lightweight, fast, and closely tied to the physical system, enabling local autonomy and immediate reactions. In the cloud, DT intelligence is more computationally intensive and supports system-level understanding, coordination, and explanation. These capabilities are complementary rather than redundant.

The overall proposed architecture simplifies the engineering of complex DT systems by decomposing the CPS into multiple coordinated twins, improves responsiveness by enabling local decisions, reduces communication through semantic distillation, and supports privacy-aware data sharing. At the same time, it enables advanced reasoning by combining the outputs of multiple twins with simulation and LLM-based interpretation.


%% file: sections/04_conclusion.tex
\section{Conclusion and Future Work}
\label{sec:conclusion}

This paper introduced the \textbf{Federated Cognitive Digital Twin}, a unified architecture that combines federation and cognition to support DTs in large-scale and distributed CPSs. The proposed approach addresses a key gap in current research. On the one hand, federated DT approaches improve structural decomposition and scalability, but often leave intelligence concentrated in centralized cloud components. On the other hand, cognitive DT approaches enhance semantic reasoning and explainability, but are typically designed as centralized solutions and therefore remain difficult to apply in latency-sensitive and distributed settings. 

The FCDT architecture bridges these two lines of work by explicitly distributing DT capabilities across the edge-to-cloud continuum. In the proposed design, local twins operate close to the physical subsystems and provide real-time monitoring, analysis, first-level decision-making, and lightweight cognitive processing. Global twins operate in the cloud and support system-level reasoning, simulation, explanation, and coordination across multiple subsystems. Their interaction through a shared data space enables loose coupling, scalable integration, and bidirectional knowledge exchange. In this way, the architecture combines local autonomy and responsiveness with global semantic reasoning and coordinated decision support.

Overall, the proposed architecture contributes a coherent way to rethink DT engineering in distributed environments. It supports the decomposition of complex CPSs into multiple coordinated twins, reduces the drawbacks of centralized architectures, and extends DTs beyond data processing toward more meaningful and explainable decision support. For these reasons, we believe that FCDT represents a promising architectural direction for DTs in scenarios such as smart cities and other complex distributed systems.

As future work, we plan to refine the architecture along both engineering and evaluation dimensions. First, we will further formalize the responsibilities, interfaces, and interaction patterns of local and global twins, including the role of the shared data space and the coordination mechanisms between edge and cloud components. Second, we will investigate model allocation strategies for deciding which cognitive capabilities should remain at the edge and which should be delegated to the cloud, considering trade-offs among latency, bandwidth, computational cost, privacy, and quality of reasoning. Third, we plan to develop a proof-of-concept implementation and assess the architecture in realistic distributed scenarios, with particular attention to responsiveness, resilience, communication overhead, and the usefulness of semantic distillation and LLM-based reasoning for decision support. 

In addition, future work should examine how the proposed architecture can support stronger interoperability and governance. This includes studying semantic integration mechanisms across heterogeneous twins, privacy-aware information sharing, and explainability techniques that make system-level recommendations more transparent to human operators. Finally, we aim to investigate tool support and model-driven engineering techniques for simplifying the design, deployment, and evolution of FCDT-based systems, so that the proposed architecture can be adopted more systematically in practice.

%% file: bibliography.bib
@inproceedings{bucaioni2025multi,
  title={Multi-partner project: A model-driven engineering framework for federated digital twins of industrial systems (matisse)},
  author={Bucaioni, Alessio and Eramo, Romina and Berardinelli, Luca and Bruneliere, Hugo and Combemale, Benoit and Khelladi, Djamel Eddine and Muttillo, Vittoriano and Sadovykh, Andrey and Wimmer, Manuel},
  booktitle={2025 Design, Automation \& Test in Europe Conference (DATE)},
  pages={1--6},
  year={2025},
  organization={IEEE}
}

@inproceedings{ferko2012supporting,
  title={Supporting technical adaptation and implementation of digital twins in manufacturing},
  author={Ferko, Enxhi and Bucaioni, Alessio and Behnam, Moris},
  booktitle={International Conference on Information Technology-New Generations},
  pages={181--189},
  year={2012},
  organization={Springer}
}

@inproceedings{ferko2023analysing,
  title={Analysing interoperability in digital twin software architectures for manufacturing},
  author={Ferko, Enxhi and Bucaioni, Alessio and Pelliccione, Patrizio and Behnam, Moris},
  booktitle={European conference on software architecture},
  pages={170--188},
  year={2023},
  organization={Springer}
}

@inproceedings{ferko2023standardisation,
  title={Standardisation in digital twin architectures in manufacturing},
  author={Ferko, Enxhi and Bucaioni, Alessio and Pelliccione, Patrizio and Behnam, Moris},
  booktitle={2023 IEEE 20th International Conference on Software Architecture (ICSA)},
  pages={70--81},
  year={2023},
  organization={IEEE}
}

@article{somma2025model,
  title={A model-driven approach for engineering Mobility Digital Twins: The Bologna case study},
  author={Somma, Alessandra and Amalfitano, Domenico and Bucaioni, Alessio and De Benedictis, Alessandra},
  journal={Information and Software Technology},
  pages={107863},
  year={2025},
  publisher={Elsevier}
}

@ARTICLE{DeBenedictis_DTHealth_2023,
  author={De Benedictis, Alessandra and Mazzocca, Nicola and Somma, Alessandra and Strigaro, Carmine},
  journal={IEEE Journal of Biomedical and Health Informatics}, 
  title={Digital Twins in Healthcare: An Architectural Proposal and Its Application in a Social Distancing Case Study}, 
  year={2023},
  volume={27},
  number={10},
  pages={5143-5154},
  doi={10.1109/JBHI.2022.3205506}}

@ARTICLE{Ferko_Architecting_2022,
  author={Ferko, Enxhi and Bucaioni, Alessio and Behnam, Moris},
  journal={IEEE Access}, 
  title={Architecting Digital Twins}, 
  year={2022},
  volume={10},
  number={},
  pages={50335-50350},
  doi={10.1109/ACCESS.2022.3172964}}

@article{Tao_DTModeling_2022,
title = {Digital twin modeling},
journal = {Journal of Manufacturing Systems},
volume = {64},
pages = {372-389},
year = {2022},
issn = {0278-6125},
doi = {10.1016/j.jmsy.2022.06.015},
author = {Fei Tao and Bin Xiao and Qinglin Qi and Jiangfeng Cheng and Ping Ji},
}

@ARTICLE{Minera_DTIoT_2020,
  author={Minerva, Roberto and Lee, Gyu Myoung and Crespi, Noël},
  journal={Proceedings of the IEEE}, 
  title={Digital Twin in the IoT Context: A Survey on Technical Features, Scenarios, and Architectural Models}, 
  year={2020},
  volume={108},
  number={10},
  pages={1785-1824},
  doi={10.1109/JPROC.2020.2998530}}

@ARTICLE{Duran_DTaaS_2026,
  author={Duran, Kubra and Verda Cakir, Lal and Yigit, Yagmur and Huseynov, Khayal and Ram Kusu, Sushmitha and Ali Ertürk, Mehmet and Canberk, Berk},
  journal={IEEE Communications Surveys \& Tutorials}, 
  title={Toward Digital Twin-as-a-Service (DTaaS) Platforms: A Survey on Architecture, Design Requirements, and Performance Metrics}, 
  year={2026},
  volume={28},
  number={},
  pages={1845-1878},
  doi={10.1109/COMST.2025.3635582}}

@article{SOMMA2026_twinarch,
title = {TwinArch: A digital twin reference architecture},
journal = {Journal of Systems and Software},
volume = {231},
pages = {112613},
year = {2026},
issn = {0164-1212},
doi = {/10.1016/j.jss.2025.112613},
author = {Alessandra Somma and Domenico Amalfitano and Alessandra {De Benedictis} and Patrizio Pelliccione},
}

@ARTICLE{Shi_EdgeComputing_2016,
  author={Shi, Weisong and Cao, Jie and Zhang, Quan and Li, Youhuizi and Xu, Lanyu},
  journal={IEEE Internet of Things Journal}, 
  title={Edge Computing: Vision and Challenges}, 
  year={2016},
  volume={3},
  number={5},
  pages={637-646},
  doi={10.1109/JIOT.2016.2579198}}

@ARTICLE{Satyanarayanan_EdgeComputing_2017,
  author={Satyanarayanan, Mahadev},
  journal={Computer}, 
  title={The Emergence of Edge Computing}, 
  year={2017},
  volume={50},
  number={1},
  pages={30-39},
  doi={10.1109/MC.2017.9}}

@INPROCEEDINGS{Pasupulati_CentralizedArchitectures_2016,
  author={Pasupulati, Renuka Prasad and Shropshire, Jordan},
  booktitle={SoutheastCon 2016}, 
  title={Analysis of centralized and decentralized cloud architectures}, 
  year={2016},
  volume={},
  number={},
  pages={1-7},
  doi={10.1109/SECON.2016.7506680}}

@INPROCEEDINGS{Vergara_FDT_2023,
  author={Vergara, Christian and Bahsoon, Rami and Theodoropoulos, Georgios and Yanez, Wendy and Tziritas, Nikos},
  booktitle={2023 IEEE/ACM 27th International Symposium on Distributed Simulation and Real Time Applications (DS-RT)}, 
  title={Federated Digital Twin}, 
  year={2023},
  volume={},
  number={},
  pages={115-116},

  doi={10.1109/DS-RT58998.2023.00027}}

@inproceedings{Vergara_FDT_2024,
author = {Vergara, Christian Roberto and Theodoropoulos, Georgios and Bahsoon, Rami and Yanez, Wendy and Tziritas, Nikos},
title = {Federated Digital Twins as an Enabling Technology for Collaborative Decision-Making},
year = {2024},
isbn = {9798400703638},
publisher = {Association for Computing Machinery},
doi = {10.1145/3615979.3662152},
booktitle = {Proceedings of the 38th ACM SIGSIM Conference on Principles of Advanced Discrete Simulation},
pages = {67–68},
numpages = {2},
series = {SIGSIM-PADS '24}
}

@Article{Papacharalampopoulos_FDT_2024,
AUTHOR = {Papacharalampopoulos, Alexios and Christopoulos, Dionysios and Karagianni, Olga Maria and Stavropoulos, Panagiotis},
TITLE = {Federation in Digital Twins and Knowledge Transfer: Modeling Limitations and Enhancement},
JOURNAL = {Machines},
VOLUME = {12},
YEAR = {2024},
NUMBER = {10},
ARTICLE-NUMBER = {701},
DOI = {10.3390/machines12100701}
}

@InProceedings{Vergara_ApproachFDT_2025,
author="Vergara-Marcillo, Christian
and Bahsoon, Rami
and Tziritas, Nikos
and Theodoropoulos, Georgios",
title="A Connectionist Approach to Federated Digital Twins",
booktitle="Computational Science -- ICCS 2025",
year="2025",
publisher="Springer Nature Switzerland",
address="Cham",
pages="60--74",
}

@article{Michael_MDE4DT_2025,
author = {Michael, Judith and Cleophas, Loek and Zschaler, Steffen and Clark, Tony and Combemale, Benoit and Godfrey, Thomas and Khelladi, Djamel Eddine and Kulkarni, Vinay and Lehner, Daniel and Rumpe, Bernhard and Wimmer, Manuel and Wortmann, Andreas and Ali, Shaukat and Barn, Balbir and Barosan, Ion and Bencomo, Nelly and Bordeleau, Francis and Grossmann, Georg and Karsai, Gabor and Kopp, Oliver and Mitschang, Bernhard and Muñoz Ariza, Paula and Pierantonio, Alfonso and Polack, Fiona A. C. and Riebisch, Matthias and Schlingloff, Holger and Stumptner, Markus and Vallecillo, Antonio and van den Brand, Mark and Vangheluwe, Hans},
title = {Model-Driven Engineering for Digital Twins: Opportunities and Challenges},
journal = {Systems Engineering},
volume = {28},
number = {5},
pages = {659-670},
doi = {10.1002/sys.21815},
year = {2025}
}

@article{Manzi_interTwin_2026,
title = {interTwin: Advancing Scientific Digital Twins through AI, Federated Computing and Data},
journal = {Future Generation Computer Systems},
volume = {179},
pages = {108312},
year = {2026},
issn = {0167-739X},
doi = {10.1016/j.future.2025.108312},
author = {Andrea Manzi and Raul Bardaji and Ivan Rodero and Germán Moltó and Sandro Fiore and Isabel Campos and Donatello Elia and Francesco Sarandrea and A. Paul Millar and Daniele Spiga and Matteo Bunino and Gabriele Accarino and Lorenzo Asprea and Samuel Bernardo and Miguel Caballer and Charis Chatzikyriakou and Diego Ciangottini and Michele Claus and Andrea Cristofori and Davide Donno and Emanuele Donno and Iacopo Ferrario and Massimiliano Fronza and Alexander Jacob and Javad Komijani and Marina Krstic Marinkovic and Federica Legger and Ivan Palomo and Estíbaliz Parcero and Rakesh Sarma and Gaurav {Sinha Ray} and Sara Vallero and Juraj Zvolensky},
}

@article{Liu_CDT_2025,
title = {A survey of cognitive digital twin and the potential use of LLMs},
journal = {Manufacturing Letters},
volume = {44},
pages = {1242-1253},
year = {2025},
note = {53rd SME North American Manufacturing Research Conference (NAMRC 53)},
doi = {10.1016/j.mfglet.2025.06.144},
author = {Yangyang Liu and Tang Ji and Xiangyu Guo and Xun Xu and Jan Polzer},
}

@article{Zheng_EmergenceCDT_2022,
author = {Xiaochen Zheng and Jinzhi Lu and Dimitris Kiritsis},
title = {The emergence of cognitive digital twin: vision, challenges and opportunities},
journal = {International Journal of Production Research},
volume = {60},
number = {24},
pages = {7610--7632},
year = {2022},
publisher = {Taylor \& Francis},
doi = {10.1080/00207543.2021.2014591},
}

@INPROCEEDINGS{Abburu_CogniTwin_2020,
  author={Abburu, Sailesh and Berre, Arne J. and Jacoby, Michael and Roman, Dumitru and Stojanovic, Ljiljana and Stojanovic, Nenad},
  booktitle={2020 IEEE International Conference on Engineering, Technology and Innovation (ICE/ITMC)}, 
  title={COGNITWIN – Hybrid and Cognitive Digital Twins for the Process Industry}, 
  year={2020},
  volume={},
  number={},
  pages={1-8},
  doi={10.1109/ICE/ITMC49519.2020.9198403}}

@INPROCEEDINGS{Zhang_EngineeringCDT_2020,
  author={Zhang, Nan and Bahsoon, Rami and Theodoropoulos, Georgios},
  booktitle={2020 IEEE International Conference on Systems, Man, and Cybernetics (SMC)}, 
  title={Towards Engineering Cognitive Digital Twins with Self-Awareness}, 
  year={2020},
  volume={},
  number={},
  pages={3891-3891},
  doi={10.1109/SMC42975.2020.9283357}}

@ARTICLE{Zhu_CDTLLM_2026,
  author={Zhu, Fang and Chen, Jiayuan and Wen, Junjie and Yang, Yuye and Yi, Changyan and Tie, Yun and Zhang, Peng and Cai, Jun and Niyato, Dusit and Guizani, Mohsen},
  journal={IEEE Communications Surveys \& Tutorials}, 
  title={From Data Mirror to Smart Copilot: A Survey on NextG Semantic Communication for Propelling Digital Twin World Into Cognitive Stage}, 
  year={2026},
  volume={28},
  number={},
  pages={4915-4947},
  doi={10.1109/COMST.2026.3665395}}

@INPROCEEDINGS{Xia_DTLLMArchitecture_2025,
  author={Xia, Yuchen and Jazdi, Nasser and Weyrich, Michael},
  booktitle={2025 IEEE 30th International Conference on Emerging Technologies and Factory Automation (ETFA)}, 
  title={An Architecture for Integrating Large Language Models with Digital Twins and Automation Systems}, 
  year={2025},
  volume={},
  number={},
  pages={1-8},
  doi={10.1109/ETFA65518.2025.11205636}}

@article{Singh_LLMFederation_2026, 
title={Learning to Collaborate: An Orchestrated-Decentralized Framework for Peer-to-Peer LLM Federation}, 
volume={40},
DOI={10.1609/aaai.v40i30.39742},
number={30}, 
journal={Proceedings of the AAAI Conference on Artificial Intelligence}, 
author={Singh, Inderjeet and Vissol-Gaudin, Eleonore and Otung, Andikan and Sekiya, Motoyoshi}, 
year={2026}, 
month={Mar.}, 
pages={25472-25480} }

@ARTICLE{Friha_LLMEdge_2024,
  author={Friha, Othmane and Amine Ferrag, Mohamed and Kantarci, Burak and Cakmak, Burak and Ozgun, Arda and Ghoualmi-Zine, Nassira},
  journal={IEEE Open Journal of the Communications Society}, 
  title={LLM-Based Edge Intelligence: A Comprehensive Survey on Architectures, Applications, Security and Trustworthiness}, 
  year={2024},
  volume={5},
  number={},
  pages={5799-5856},
  doi={10.1109/OJCOMS.2024.3456549}}

@INPROCEEDINGS{Cai_EdgeLLM_2024,
  author={Cai, Fenglong and Yuan, Dong and Yang, Zhe and Cui, Lizhen},
  booktitle={2024 IEEE International Conference on Web Services (ICWS)}, 
  title={Edge-LLM: A Collaborative Framework for Large Language Model Serving in Edge Computing}, 
  year={2024},
  volume={},
  number={},
  pages={799-809},
  doi={10.1109/ICWS62655.2024.00099}}

@inproceedings{Rao_ECOLLM_2024,
author = {Rao, Kunal and Coviello, Giuseppe and Benedetti, Priscilla and Giuseppe De Vita, Ciro and Mellone, Gennaro and Chakradhar, Srimat},
title = {ECO-LLM: LLM-based Edge Cloud Optimization},
year = {2024},
publisher = {Association for Computing Machinery},
doi = {10.1145/3660605.3660941},
booktitle = {Proceedings of the 2024 Workshop on AI For Systems},
pages = {7–12},
numpages = {6},
location = {Pisa, Italy},
series = {AI4Sys '24}
}

@inproceedings{Hao_SLM_LLM_2024,
author = {Hao, Zixu and Jiang, Huiqiang and Jiang, Shiqi and Ren, Ju and Cao, Ting},
title = {Hybrid SLM and LLM for Edge-Cloud Collaborative Inference},
year = {2024},
publisher = {Association for Computing Machinery},
address = {New York, NY, USA},
doi = {10.1145/3662006.3662067},
booktitle = {Proceedings of the Workshop on Edge and Mobile Foundation Models},
pages = {36–41},
numpages = {6},
location = {Minato-ku, Tokyo, Japan},
series = {EdgeFM '24}
}

@article{Rosendo_EdgeCloudContinuum_2022,
title = {Distributed intelligence on the Edge-to-Cloud Continuum: A systematic literature review},
journal = {Journal of Parallel and Distributed Computing},
volume = {166},
pages = {71-94},
year = {2022},
doi = {10.1016/j.jpdc.2022.04.004},
author = {Daniel Rosendo and Alexandru Costan and Patrick Valduriez and Gabriel Antoniu},
}
